\documentclass[aps,prl,twocolumn,showpacs,groupedaddress]{revtex4} 
\usepackage{graphicx}  % needed for figures
\usepackage{dcolumn}   % needed for some tables
\usepackage{bm}        % for math
\usepackage{amssymb}   % for math
\usepackage{color}	   % for text highlighting
\usepackage{amsmath}  
\begin{document}

\title{Growth modes of quasicrystals}

\author{C. V. Achim}
\affiliation{Institut f\" ur Theoretische Physik II: Weiche Materie, Heinrich-Heine-Universit\" at D\" usseldorf,  D-40204 D\" usseldorf, Germany} 
\author{M. Schmiedeberg}
\email[E-Mail: ]{schmiedeberg@thphy.uni-duesseldorf.de}
\affiliation{Institut f\" ur Theoretische Physik II: Weiche Materie, Heinrich-Heine-Universit\" at D\" usseldorf,  D-40204 D\" usseldorf, Germany} 
\author{H. L\"owen}
\affiliation{Institut f\" ur Theoretische Physik II: Weiche Materie, Heinrich-Heine-Universit\" at D\" usseldorf,  D-40204 D\" usseldorf, Germany}

\begin{abstract}
The growth of quasicrystals, i.e., aperiodic structures with long-range 
order, seeded from the melt is investigated using
a dynamical phase field crystal model. Depending on the thermodynamic conditions,
two different  growth modes 
are detected, namely defect-free growth of the stable quasicrystal and a
mode dominated by phasonic flips which are incorporated as local defects into the grown structure
such that random tiling-like ordering emerges. 
The latter growth mode is unique to quasicrystals and can be verified 
in experiments on one-component mesoscopic systems.
\end{abstract} 
%82.70.Dd:colloids 
%61.44.Br:Quasicrystals
%64.75.Yz Self-assembly
%64.70.D- Solid-liquid transitions
%47.54.-r Pattern selection; pattern formation
%61.50.Ah Theory of crystal structure, crystal symmetry
%05.20.-y Classical statistical mechanics
%64.60.Cn Order-disorder transformations
%81.10.Aj Theory and models of crystal growth

\maketitle

Quasicrystals are aperiodic structures that possess long range positional and orientational order \cite{shechtman1984,levineprl1984}. Since their discovery by Shechtman \cite{shechtman1984}, hundreds of quasicrystals have been reported and confirmed. Most of them are metallic alloys (see, e.g., \cite{steurerchemrev2012,macia2006}) but more recently they
have also beend found  in soft-matter systems that are made, e.g., by amphiphilic molecules \cite{fischer2011}, supramolecular dendritic systems \cite{zengnat2004,zengcuropin2005}, or by star block copolymers \cite{takano2005,hayashidaprl2007}. 
Such soft matter quasicrystals can provide scaffolds for photonic materials \cite{lifshitzbook2009} and serve as
well-characterized mesoporous matrices \cite{percec2009,smithpolsci2005}. In general, quasicrystals occur 
either as defect-free structures stabilized 
by energy \cite{strandburgprl1989,tangprb1990,tangprl1990,shawprb1991,jeongprb1993} or as 
locally disordered phases, leading to random tiling like structures, 
stabilized by entropy~\cite{trebinprl2012}. 

One of the key issues for quasicrystal formation is to understand their growth mechanism out of an undercooled melt.
Unlike ordinary growth of periodic crystals where a layer-by-layer mode is possible, quasicrystals
lack any strict sequential growth mode due to their aperiodicity which renders their formation quite complex.
Based on atomistic simulations, it has been proposed that instead first clusters are formed in the fluid which then assemble
in the growing solid-fluid interface \cite{keysprl2006} but the fundamentals and details for quasicrystal growth
are far from being 
understood. In particular, the incorporation of defects into the emerging structure
during the growth process plays the leading role to discriminate
 between grown defect-free and random-tiling-like 
quasicrystals. 

In this letter we explore the growth behavior of quasicrystals using
an appropriate dynamical phase field crystal model  with two incommensurate length scales
which exhibits stable defect-free quasicrystals in equilibrium.
Depending on the thermodynamic conditions (such as undercooling and distance from the triple point),
we find two different growth regimes for quasicrystals. There is either
 a defect-free growth into the stable quasicrystal or a
mode dominated by phasonic flips which are incorporated as local defects into the grown structure
such that a metastable random tiling-like ordering emerges. 
The latter growth mode is unique to quasicrystals and can be verified 
in experiments on one component mesoscopic systems which exhibit quasicrystalline order. 
Our findings do not only provide a microscopic (i.e. particle-resolved) understanding of the
growth processes on the scale of the particle motion but can also be exploited to steer
the emerging quasicrystalline texture by the thermodynamic conditions.

Phase field crystal (PFC) models in general employ a coarse-grained free-energy expansion with respect to a scalar density field $\psi(\vec r,t)$, which is related to the one-particle density $\rho(\vec r)$ \cite{jaatinenpre2010}. In a fluid phase, the density field $\psi(\vec r,t)$ is constant, while crystal phases are described by periodic or quasi-periodic modulations of $\psi(\vec r,t)$. The PFC formalism was introduced first in the framework of periodic pattern formation as a Swift-Hohenberg  free-energy \cite{swiftpra1977,crossbook2009} and was later derived from microscopic theories \cite{loewenpre2009,elderprovatasprb}. In the original form, only one preferred length scale is regarded in the free energy, which in two dimensions allows for only two periodic solid phases consisting of triangular order or stripes \cite{elderprl2002,elderpre2004}. By adding a second length scale, quasicrystalline structures can be found that minimize the free energy \cite{lifshitzprl1997,rottlerjphycondmatt2012}). Therefore this PFC model describes one-component systems whose particles interact according to a pair potential with at least two lengths scales \cite{lifshitzprl1997,lifshitzprb2011}. Quasicrystals were predicted in such systems theoretically \cite{denton1988,engelprl2007,archerprl2013,barkan2014,dotera2014} and one-component soft matter quasicrystals \cite{fischer2011,zengnat2004,zengcuropin2005,takano2005,hayashidaprl2007} are expected to occur due to such pair interactions \cite{lifshitzprb2011}.

In our dynamical PFC model, the density field will evolve towards a local minimum following conserved dynamics \cite{elderprl2002}, i.e.,
\begin{equation}\label{eq:dPFC}
 \frac{\partial \psi(\vec r,t)}{\partial t}=\nabla^2\left[\frac{\delta F[\psi(\vec r, t)]}{\delta \psi(\vec r,t)}\right],
\end{equation}
where we employ an expansion of the free energy with two length scales \cite{rottlerjphycondmatt2012,anmerkung}
\begin{eqnarray}
 F[\psi(\vec r)]=\int d\vec r&&\left[\frac{1}{2}\psi(\vec r)\left\{-\epsilon+\prod_{j=1}^2\left(k_j^2+\nabla^2\right)^2\right\}\psi(\vec r) \right. \nonumber\\
&&\left.\ \ +\frac{1}{4}\psi(\vec r)^4\right].\label{eq:free-energy}
\end{eqnarray}
We choose the long length scale $l_1=2\pi/k_1$ such that $k_1=1$ and the diffusion constant is set to one such that the time is given in units of the time an individual particles need to diffuse a length $l_1/\pi$. The remaining parameters of this model are the ratio of the length scales given by $k_2$, the mean density $\bar\psi$, and $\epsilon$, which can be interpreted as the mean field temperature \cite{elderprl2002,elderpre2004}. Quasicrystalline structures that minimize the free energy can be found if the length scales are properly adjusted \cite{lifshitzprl1997,lifshitzprb2011,rottlerjphycondmatt2012,barkan2014}. Here we consider the cases of 12- and 10-fold symmetry 
 by using $k_2=2\cos(\pi/12)$ or $k_2=2\cos(\pi/5)$, respectively \cite{rottlerjphycondmatt2012,barkan2014}. 

\begin{figure}
\includegraphics[width=\columnwidth]{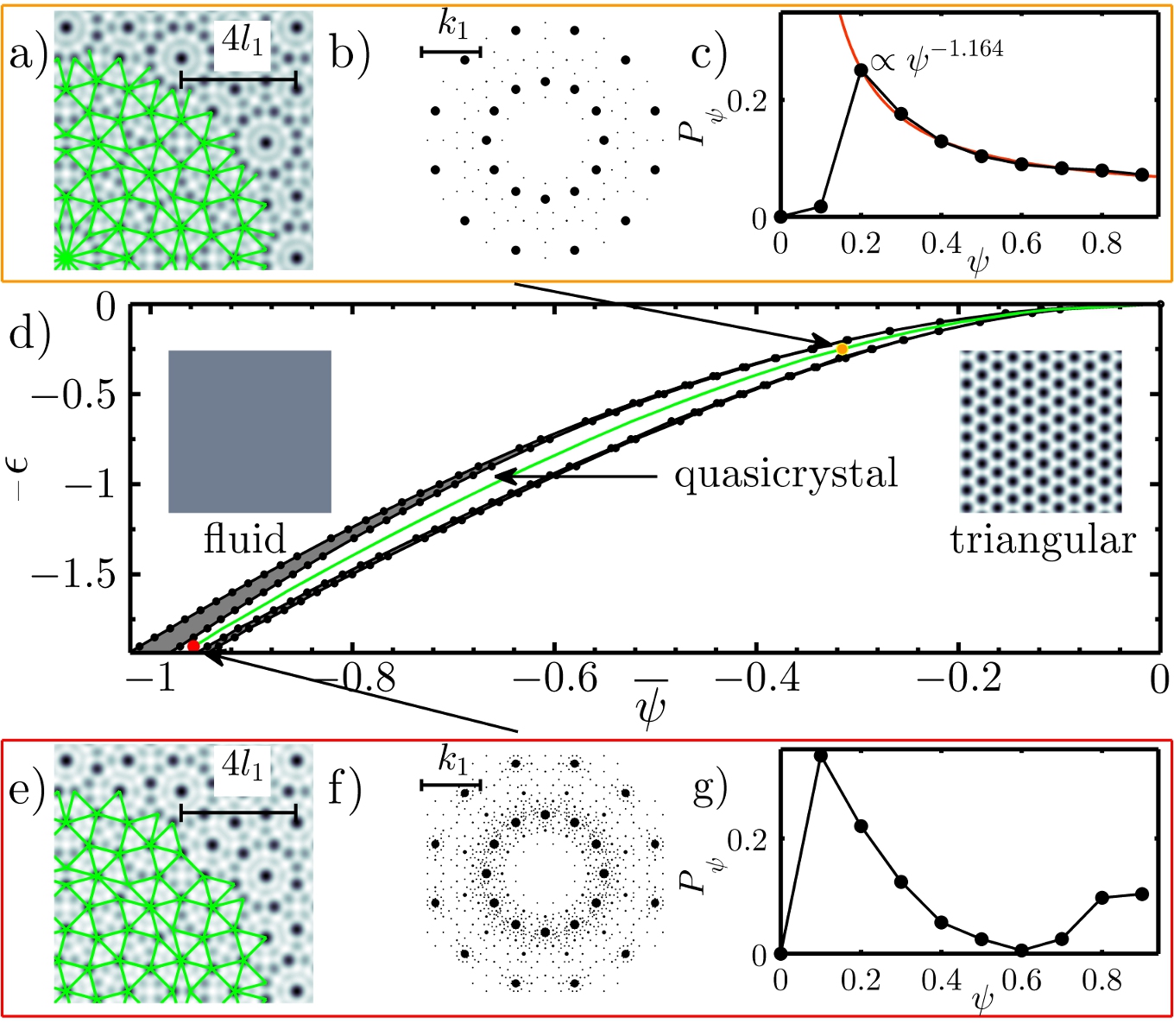}
\caption{\label{fig:phdi} Equilibrium phase behavior determined with the PFC model for $k_2=2\cos(\pi/12)$. In the center (d) the phase diagram with stable triangular, 12-fold quasicrystalline, and fluid phases is shown depending on the reduced temperature $\epsilon$ and the reduced average density $\bar \psi$. Coexistence regions are shaded in grey. The green line denotes the path considered in Fig. \ref{fig:flips}(g). (a,e) Quasicrystalline density fields with possible tilings, (b,f) the corresponding structure factors, and (c,g) the distributions of heights of local maxima of the density field (a-c) close to the triple point for $\epsilon=0.25$ and $\bar \psi=-0.314$ (orange point in (d)) and (e-g) far away from the triple point for $\epsilon=1.90$ and $\bar \psi=-0.947$ (red point). The red line in (c) is a power law fit.
}
\end{figure}

In Figure \ref{fig:phdi}(d)  the equilibrium phase diagram is shown in the plane spanned by the triple point distance 
 $\epsilon$ and the mean density $\bar \psi$ for the case $k_2=2\cos(\pi/12)$, i.e., the length scales are chosen to support 12-fold quasicrystalline order. Note that the diagram is symmetric with respect to $\bar \psi=0$ and therefore we only plot negative values of $\bar \psi$. Furthermore, for $\epsilon>0$ only a fluid phase occurs. The phase diagram was constructed by comparing the free energies of the fluid, a quasicrystal with 12-fold rotational symmetry, triangular phases, and stripe phases. We find stable fluid and quasicrystalline phases as well as triangular order, whose lattice constant is given by the small length scale. The phase transitions are of first order and we employ a common tangent construction in order to obtained the coexistence regions marked by gray color in Fig. \ref{fig:phdi}(d).

At $\bar \psi=0$ and $\epsilon=0$, there is a triple point where the stable quasicrystalline phase with 12-fold rotational symmetry, triangular ordering, and the fluid phase meet. Interestingly, the quasicrystalline phase looks different depending  on the distance $\epsilon$ from the triple point. For small $\epsilon$, we observe that every local symmetry center, which in the following we will call flower in accordance with \cite{bechingersoftmatter2013}, is surrounded by twelve density peaks of approximately similar height (Fig. \ref{fig:phdi}(a)). However, for large $\epsilon$ (Fig. \ref{fig:phdi}(e)), the density peaks in a flower usually have very different heights. Furthermore, in the structure factor of a quasicrystal close to the triple point, the height of the main Bragg peaks is significantly larger than that of the side peaks (Fig. \ref{fig:phdi}(b)) while for a quasicrystal far away from the triple point side peaks are more pronounced (Fig. \ref{fig:phdi}(f)). Finally, we also analyzed the distribution of heights of local maxima of the density field. Close to the triple point, we find a power law distribution (Fig. \ref{fig:phdi}(c)) which is in agreement to what one expects from the sum of plane waves or the pattern created by interfering laser beams (cf. \cite{schmiedebergepje2007}). Far away from the triple point, the distribution of the height of local maxima exhibits a minimum for maxima with intermediate height (Fig. \ref{fig:phdi}(g)) indicating that there are mainly low or high peaks in the density field. 

\begin{figure*}
\includegraphics[width=1.99\columnwidth]{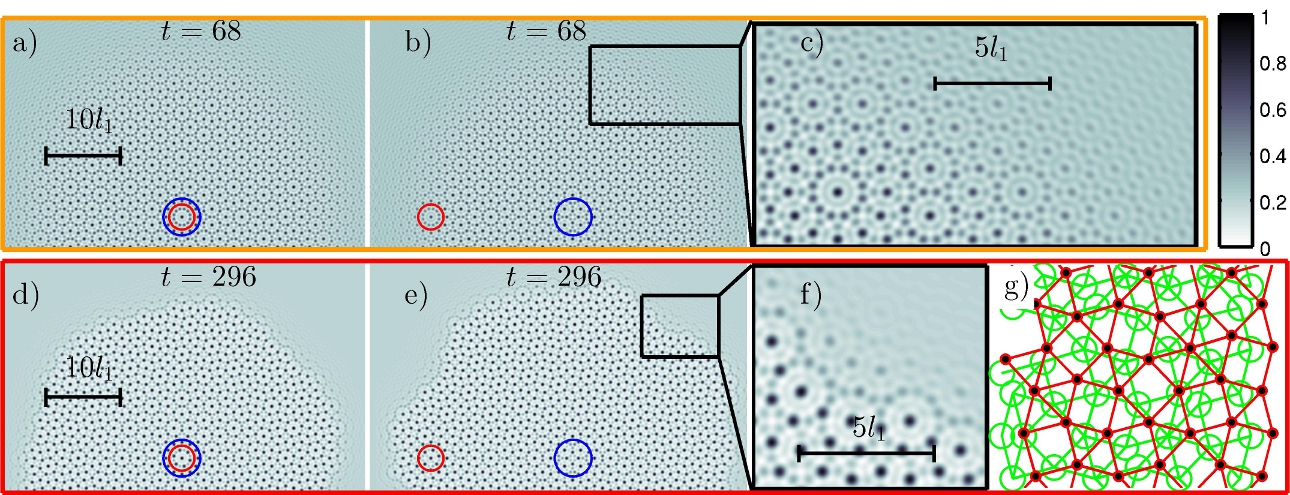}
\caption{\label{fig:smepsgr} Snapshots of structures during the growth. The plotted density is scaled and shifted such that its values in bulk range from 1 for the maximum (depicted black) to 0 for the minimum (white). The blue circles mark the positions of the original seed, which either is (a,d) around the global symmetry center (red circle) or (b,e) at another position. The parameter sets correspond to the orange and red points in Fig. \ref{fig:phdi}(d), i.e., (a-c) close to the triple point leading to a perfect quasicrystal or (d-g) far away from the triple point such that a random tiling-like phase develops. (c,f) Enlargements of the growth fronts in (b,e). (g) Comparison of the grown structure (red tiling) and the perfect equilibrium quasicrystal (green tiling). Movies of (a), (b), (d), and (e) are available in the supplemental materials \cite{suppl}. 
}
\end{figure*}

In Fig. \ref{fig:smepsgr} we present the growth behavior of a quasicrystal. Our calculations are started with a quasicrystalline seed that is placed into a supercooled fluid surrounding and prescribes the initial density field. 
The seeds (marked by blue circles)  are circular 
cut-outs from the equilibrium quasicrystal either around the global symmetry center (Figs. \ref{fig:smepsgr}(a),(d)) or at another position (Figs. \ref{fig:smepsgr}(b),(e)). For clarity, the global symmetry center, which is the only point in the quasicrystal with perfect rotational symmetry, is marked by red circles. 

For the parameters marked by the orange point in the phase diagram of Fig. \ref{fig:phdi}(d) close to the triple point, we observe a very broad growth front that spans over more than ten times the small length scale $l_1$ (Figs. \ref{fig:smepsgr}(a)-(c)). We checked different seed shapes and sizes (not shown) as well as different seed positions (see, e.g., Fig. \ref{fig:smepsgr}(a,b)). Below a critical seed size, which encloses about 400 density maxima, the system falls back
towards the  metastable fluid. Conversely, above this critical seed size, there is growth and the emerging
structure does not depend 
on the details of the seed. Even the global symmetry center is reconstructed perfectly no matter where the growth process is seeded (cf. \ref{fig:smepsgr}(b)). Therefore, interestingly, all the information about the complete quasicrystal 
is encoded in the seed for this case.

In contrast, the growth depicted in Figs. \ref{fig:smepsgr}(d)-(g) for the parameters indicated by the red point in Fig. \ref{fig:phdi}(d), i.e., far away from the triple point, leads to metastable configurations that depend on size, position, and shape of the initial seed (see, e.g. Figs. \ref{fig:smepsgr}(d),(e)). Furthermore, the growth front does not propagate uniformly but changes between a fast progression and almost stuck stages (see also movies in the supplemental materials \cite{suppl}). The grown structures possess a lot of local defects. For example, in Fig. \ref{fig:smepsgr}(g) we show a comparison of a tiling of the grown structure (red) and the structure of the equilibrium quasicrystal (green). The global symmetry center usually is not reconstructed (see, e.g., Figs. \ref{fig:smepsgr}(e)). Note that though there are a lot of local defects, we rarely observed global dislocations (using the methods described in \cite{lifshitzapcryst2013}, see also supplemental materials \cite{suppl2}). Therefore the random tiling-like growth mode differs significantly from the well-studied situation in periodic crystals, where the growth especially in the coexistence region might lead to structures with numerous dislocations \cite{toth2010,granasysoftmat}.

The width of the quasicrystal-fluid interface is about one or two times $l_1$ (Fig. \ref{fig:smepsgr}(f)). Furthermore, the interface is not smooth laterally. The interface itself is mainly composed of
complete flowers, i.e., the growing quasicrystal completes flowers even if at the respective position there should be no flower in the perfect equilibrium quasicrystal.  Therefore, the grown structures possess a lot of local disorder and its overall shape usually is no longer circular but strongly faceted. 

The occurence of the different growth modes  is  probably related to the role of the long length scale $l_1$ during the interface propagation. While close to the triple points both length scales are equally important, far away from the triple point $l_1$ seems to dominate. For example, flowers, whose diameter is $2l_1$, are completed precipitately, pronounced density peaks that usually occur at a distance $l_1$ can be found more often then peaks of intermediate height (cf. Fig. \ref{fig:phdi}(g) in contrast to Fig. \ref{fig:phdi}(c)), and the tiling of a grown structure is dominated by squares and triangles with side length $l_1$ while rhombs are rare (see red tiling of the grown structure in Fig. \ref{fig:phdi}(g) in contrast to the green tiling of the equilibrium quasicrystal).

\begin{figure}
\includegraphics[width=\columnwidth]{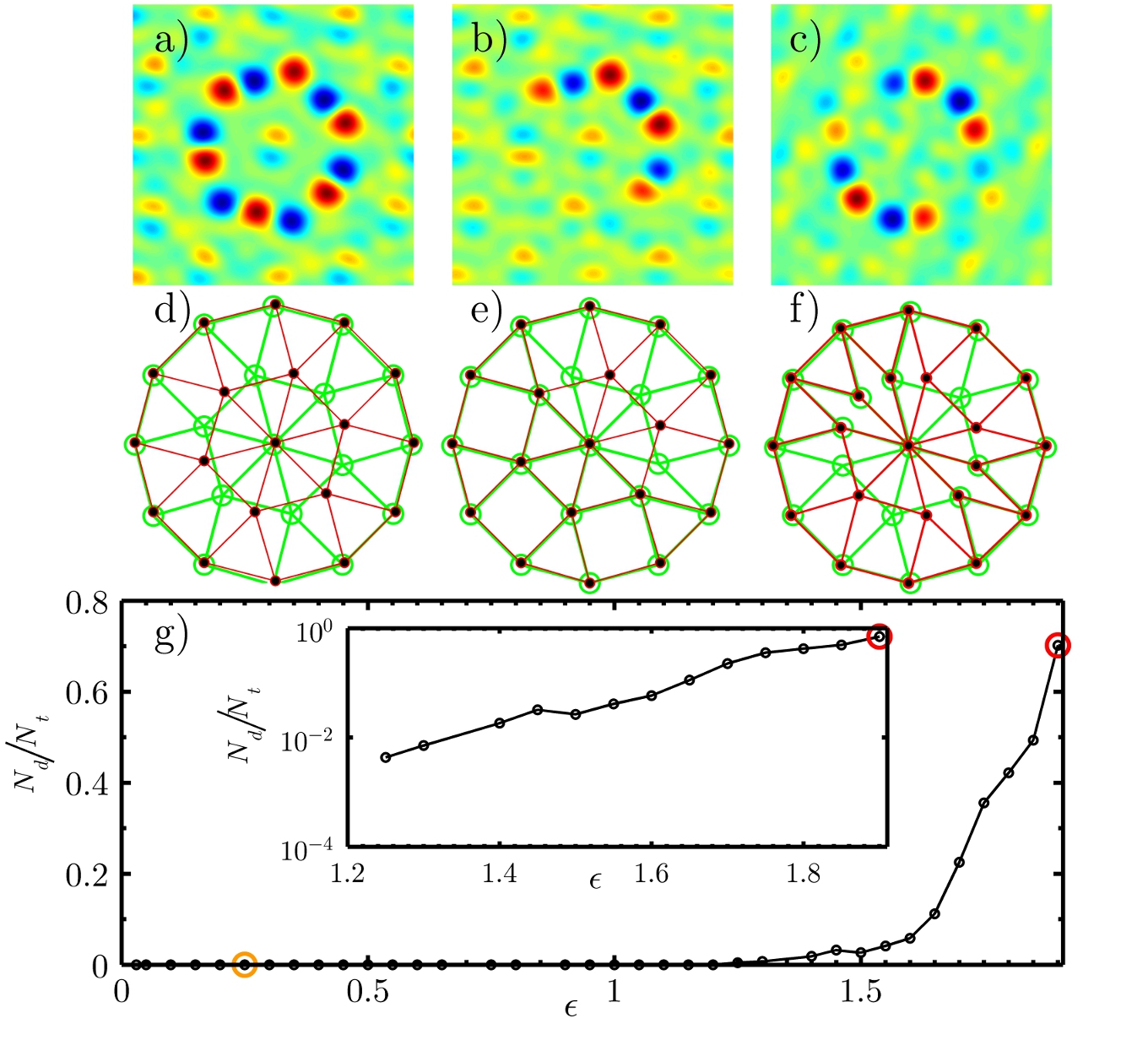}
\caption{\label{fig:flips} (a-f) Examples of phasonic flips that occur during growth for parameters far away from the triple point. (a-c) Difference of the density field of the equilibrium quasicrystal and the density field of the grown structure. (d-f) The corresponding tilings for the equilibrium (green) and the grown structure (red). (g) Number of defects $N_d$ divided by the total number of analyzed particles $N_t$ for structures grown along the green line shown in \ref{fig:phdi}(d). The inset shows the same plot in log-linear representation. The orange and the red circle mark the results for the parameters used in Fig. \ref{fig:smepsgr}. 
}
\end{figure}

In order to analyze the difference between the two growth regimes in more detail and to find the crossover between the two regimes, we compared the density field of the equilibrium quasicrystal and the density field of a structure grown from a seed containing 650 maxima around the global symmetry center for parameters along the green path shown in the phase diagram in Fig. \ref{fig:phdi}(d). We determined the places where the difference of the corresponding density fields exceeds $0.2$ of the maximum amplitude of the equilibrium density field. Such pronounced differences only occur far away from the triple point and correspond to local defects. We find that the local defects always occur in groups as depicted in Fig. \ref{fig:flips}(a)-(c). Such correlated rearrangements are known as phasonic excitations \cite{engelprl2007,steinhardtPRB34a,steinhardtPRB34,kromerprl2012,engelprb2010}. Phasons are additional degrees of freedom in quasicrystals that do not cost any free energy in the long wavelength limit \cite{steinhardtPRB34a,steinhardtPRB34} and that lead to correlated particle trajectories \cite{kromerprl2012}. Localized phasonic excitations can also be depicted as local changes of a tiling (see Fig. \ref{fig:flips}(d)-(f)) and correspond to so-called phasonic flips (cf. \cite{engelprl2007,engelprb2010}). The relative defect concentrations in structures grown at least by a radius $60 l_1$ around the perfect symmetry center are plotted in Fig. \ref{fig:flips}(g) as a function of $\epsilon$, where $\bar \psi$ is chosen such that one is on the green line depicted in Fig. \ref{fig:phdi}(d). Up to $\epsilon=1.2$ no defects occurred in the circle where we analyzed the structures. For $\epsilon\geq 1.25$ the number of defects increases almost exponentially (a log-linear plot is presented in the inset of Fig. \ref{fig:flips}(g)) until it approaches 1 for $\epsilon\approx 1.9$ where the grown configuration has a completely different structure and counting of distinct defects becomes difficult.

\begin{figure}
\includegraphics[width=\columnwidth]{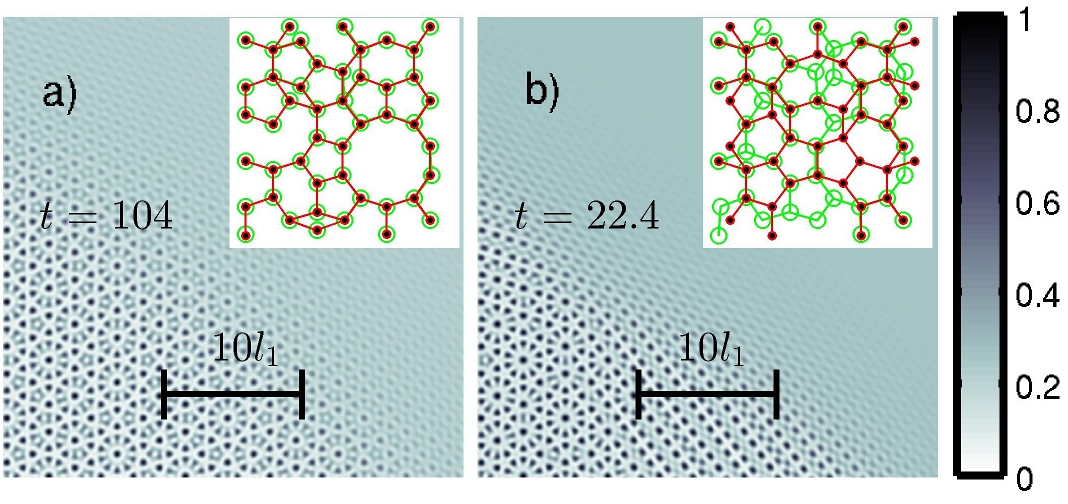}
\caption{\label{fig:otherphases} Snapshots of the growth of a quasicrystal with 10-fold symmetry. (a) Close to the triple point, for $\epsilon=0.0685$, a defect-free quasicrystal grows. (b) For $\epsilon=0.52$ the growth front is sharper and the grown structure differs from the equilibrium quasicrystal. Close to the growth front, Archimedean-like tilings are observed that are periodic in one direction and have been obtained for particles that are deposited on quasicrystalline substrates \cite{mikhael2008}. The insets show comparisons of the tilings corresponding to the equilibrium (green) and the grown structures (red). Movies are included in the supplemental materials \cite{suppl}.
}
\end{figure}

So far we have analyzed the growth of quasicrystals with $12-$fold rotational symmetry. In Fig. \ref{fig:otherphases} we demonstrate that the same two growth regimes also occur for a quasicrystal with $10$-fold symmetry where $k_2=2\cos(\pi/5)$ was chosen in Eq. (\ref{eq:free-energy}). Therefore, evidence is provided that in case of quasicrystals with two incommensurate length scales the reported growth mechanisms are independent from the specific symmetry of the quasicrystal. Concerning quasicrystals with more than two incommensurate length scales, we expect that it is harder to balance all length scales during the growth process such that defect-free growth will rarely occur. This might be an additional explanation why quasicrystals with three or more incommensurate length scales rarely occur in nature \cite{mikhaelpnas2010,schmiedebergjphyscondmatt2012,fischer2011} and are only observed as random tiling-like phases in simulations \cite{dotera2014}.

In conclusion, we presented a theory to describe the growth of quasicrystals and discovered two different growth regimes. While in one regime defect-free quasicrystals could grow, in the other regime phasonic flips were built in due to a dominating long length scale such that the final grown structures correspond to random tiling-like phases. In order to obtain defect-free growth, there has to be a balance of the growth in both length scales. For the random tiling-like growth, the growth is dominated by the long length scale. As a consequence, there is a tendency to complete flowers at the interface, which might be related to a cluster-driven growth (cf. \cite{keysprl2006}). 

Our results can be used to achieve an improved control of the growth of defect-free or random tiling-like quasicrystals in simulations or within experimental systems, e.g., the mesoscopic one-component quasicrystals mentioned in the introduction. We want to point out that it is not necessary to explore the whole phase behavior in order to determine the growth regimes. The type of growth can also be identified by analyzing the equilibrium structure (cf. Fig. \ref{fig:phdi}). In the regime with defect-free growth the Bragg peaks of the structure factor are more pronounced than in the growth regime where phasonic flips occur. Furthermore, the distribution of heights of the density maxima in the defect-free growth regime is a power law, while in the growth regime with defects this distribution possesses two maxima, one denoting shallow maxima and one for the most pronounced maxima. In the case of cluster quasicrystals, the density distribution can be measured directly by determining the distribution of cluster sizes \cite{barkan2014}. In case of other quasicrystals, it can be deduced from the statistical probability density.

\begin{acknowledgements}

We thank A. Archer, K. Barkan, A. Mijailovi\'c, and M. Sandbrink for helpful discussions. C.V.A. and H.L. received support by the ERC Advanced Grant INTERCOCOS (Grant No. 267499). M.S. was supported by the DFG within the Emmy Noether program (Schm 2657/2).

\end{acknowledgements}

\end{document}